\documentstyle[aps,preprint,floats,epsfig]{revtex}
\begin{document}
\tighten

\title{THE NUCLEON SPIN POLARIZABILITY \\ 
  AT ORDER ${\cal O}(p^4$) IN CHIRAL PERTURBATION THEORY}
\author{Xiangdong Ji, Chung-Wen Kao and Jonathan Osborne}
\bigskip

\address{
Department of Physics \\
University of Maryland \\
College Park, Maryland 20742 \\
{~}}

\date{UMD PP\#00-017 ~~~DOE/ER/40762-194~~~ August 1999}

\maketitle

\begin{abstract}
We calculate the forward spin-dependent 
photon-nucleon Compton amplitude as a function of 
photon energy at the next-to-leading 
(${\cal O}(p^4)$) order in chiral perturbation 
theory, from which we extract the contribution to 
nucleon spin polarizability. The result 
shows a large correction to the leading order
contribution. 

\end{abstract}
\pacs{xxxxxx}

\narrowtext

In recent years, the static properties of the nucleon
have been under intense theoretical and experimental
investigations. Examples include the various elastic
form factors measurable in low-energy electroweak 
processes and parton distributions accessible through 
hard scattering. The nucleon polarizabilities comprise
yet another type of observable 
which characterizes the response of the 
nucleon when exposed to external
electromagnetic fields. They can be measured 
through low-energy Compton
scattering, which has become feasible recently
thanks to novel advances in experimental 
technology. For
a summary of theoretical and experimental progress
in studying the polarizabilities, the reader can 
consult Ref. \cite{review}. 

In this letter, we study the spin
polarizability of the nucleon, which we define
here via a spin-dependent forward Compton amplitude. The photon-nucleon 
forward Compton amplitudes are related to
\begin{equation}
     T^{\mu\nu} = i\int d^4\xi e^{iq\cdot \xi}
     \langle PS|{\rm T} J^{\mu}(\xi) J^{\nu}(0)|PS\rangle \ . 
\end{equation}
where $|PS\rangle$ is the covariantly-normalized
ground state of a nucleon of momentum $P^\mu$ 
and spin polarization $S^\mu$.
$J_\mu = \sum_i e_i \bar \psi_i\gamma_\mu \psi_i$
is the electromagnetic current (with $\psi_i$ the quark field 
of flavor $i$ and $e_i$ its charge in units of the proton charge).
The four-vector $q^\mu$ is the photon 
four-momentum. Using Lorentz symmetry, parity and 
time-reversal invariance, one can express the spin-dependent 
($\mu\nu$ antisymmetric) part of $T^{\mu\nu}$ in terms 
of two scalar functions:
\begin{equation}
     T^{[\mu\nu]}(P,q,S)  = -i\epsilon^{\mu\nu\alpha\beta}
      q_\alpha \left[S_\beta S_1(\nu)
      +
    \left(M\nu\, S_\beta -S\cdot q \,P_\beta\right) S_2(\nu)\right] \ ,
\end{equation} 
where $\nu$ is the energy of the real photon and $M$ is the nucleon
mass ($\epsilon^{0123}=+1$). $S_{1,2}(\nu)$ are 
the spin-dependent, invariant Compton 
amplitudes. Of course, in real photon scattering 
$S_2(\nu)$ decouples and one can measure  
the $S_1(\nu)$ amplitude only. The relation between 
$S_1$ and the traditional amplitude 
$f_2(\nu)$ (see Ref. \cite{f2}) is 
\begin{equation}
   f_2(\nu) = {\alpha_{\rm em}\over 2} S_1(\nu) \ . 
\end{equation}
Through crossing symmetry, it is easy to 
see $S_1(\nu)$ is even in $\nu$.  
The spin-dependent polarizability $\gamma$ is
defined as 
\begin{equation}
     \gamma = \left.{df_2(\nu)\over d\nu^2}\right|_{\nu=0}\,\,\,\, ,
\end{equation}
which is just the slope of $f_2(\nu)$ at $\nu^2=0$. 

Since the spin polarizability is a low-energy observable,
it is natural to explore its physical content in 
chiral perturbation theory ($\chi$PT), or more broadly 
in low-energy effective theories. In $\chi$PT, one
considers the pion mass $m_\pi$ and the external 
three-momentum $\vec{p}$
small compared to any other scales in the problem.
In low-energy effective field theories one considers
expansions also in terms of other small parameters, 
such as the mass difference $\Delta$
between the nucleon and delta resonance. Here the 
expansion parameter is generically denoted as $\epsilon$. 
Bernard, Kaiser,
Kambor, and Meissner have studied 
the spin polarizability in $\chi$PT \cite{bernard}.
They found that at the leading-order (${\cal O}(p^3)$)\  
\begin{eqnarray}
   \gamma_{p,n}^{{\cal O}(p^3)} &=& {\alpha_{\rm em} 
    g_A^2 \over 24\pi^2 f_\pi^2 m_\pi^2} \nonumber \\
     & = & 4.4 \cdot 10^{-4}~ {\rm fm}^{4}\,\,\,\, ,
\end{eqnarray}
which diverges as 1/$m_\pi^2$ in the chiral limit.
The above result was obtained formally in the limit of the
nucleon mass going to infinity (heavy-baryon chiral
perturbation theory) \cite{heavybaryon}, a trick to make 
power counting manifest in Feynman diagrams.  

Being a spin-dependent quantity, the spin-polarizability
undoubtedly receives a large contribution from the 
delta resonance. In pure $\chi$PT, the delta resonance
contributes through counter terms and the effect appears
at ${\cal O}(p^5)$ (see the discussion below). 
However, the nucleon 
and delta mass difference is of order $m_\pi$
and vanishes in the large-$N_c$ (number of colors) limit.
Therefore it is reasonable to include the delta resonance 
explicitly
and consider an expansion with $\Delta$ considered
a small parameter. Using $\epsilon$ to denote $m_\pi$,
$\Delta$, and any small momentum, one finds 
an ${\cal O}(\epsilon^3)$ contribution from the 
tree diagram with an intermediate delta, 
\begin{eqnarray}
   \gamma_{p,n}^{{\cal O}(\epsilon^3)\rm \Delta-tree}
    & = & - {\alpha_{\rm em} \over 9}{G_1^2\over M^2\Delta^2}  
          \nonumber \\
    & = & - 2.4 \cdot 10^{-4} {\rm fm}^4 \ ,      
\end{eqnarray}
where $G_1$ is a spin-flip $N$-$\Delta$-$\gamma$
coupling and its numerical estimate of 3.85 is taken from
Ref. \cite{hemmert} (in the large $N_c$ limit, it
takes the value of $3/(2\sqrt{2})\kappa_V$). 
Not surprisingly, the above contribution is large 
and negative and almost entirely cancels the 
one-loop chiral contribution. At the same order, 
the delta resonance also contributes through
one-loop intermediate states : 
\begin{equation}
   \gamma_{p,n}^{{\cal O}(\epsilon^3)\rm \Delta-loop}
  = {\alpha_{\rm em}g_A^2\over 3\pi(4\pi f_\pi)^2}
  \left\{{2\over m_\pi^2} - {\Delta^2+2m_\pi^2\over
   (\Delta^2-m_\pi^2)^2}  
   + {3\Delta m_\pi^2\over (\Delta^2-m_\pi^2)^{5/2}}
  \ln\left[{\Delta\over m_\pi}
   + \sqrt{{\Delta^2\over m_\pi}-1}\right]\right\} \ . 
\end{equation}
With realistic parameters, the above 
contribution is much smaller than the other two 
contributions. The total ${\cal O}(\epsilon^3)$ 
result is different from that in Ref. \cite{bernard}
because of the nonrelativistic expansion and
the newly-fitted $G_1$ parameter.

In this letter, we report a calculation at 
next-to-leading order (${\cal O}(p^4)$) in 
chiral perturbation theory. The
result has a comparable size to 
that at ${\cal O}(p^3)$ and the sign is 
opposite. We have not considered 
the tree-level delta contribution 
at ${\cal O}(\epsilon^4)$, 
as it involves a number of unknown parameters on which one
has no firm theoretical handle. The delta-loop 
contribution can be evaluated at this order, 
but we expect its contribution to be small. 

Before we present the details and the result of our
calculation, it is useful to recall some of the standard
infrared power counting techniques in effective field theories
as formulated in the heavy-baryon approach. The 
full lagrangian (including nucleon, delta, photon, and pion fields)
can be expanded :
\begin{equation}
    {\cal L} = {\cal L}^{(1)} + {\cal L}^{(2)} 
     + {\cal L}^{(3)} + {\cal L}^{(4)} + ...
\end{equation}
where ${\cal L}^{(n)}$ contains terms of order $\epsilon^n$,
with one power of $\epsilon$ assigned to each derivative, pion
mass, photon field, nucleon-delta mass difference, etc.  The infrared power
of a Feynman diagram is generated from the vertices
and the propagators. For polarized Compton scattering,
Feynman diagrams start at ${\cal O}(\epsilon^3)$. 
For tree diagrams at this order, one needs to consider
vertices at ${\cal O}(\epsilon^n)$ with $n\le 3$. 
For one-loop diagrams, we need to consider 
vertices at ${\cal O}(\epsilon)$ only. In general,
at order ${\cal O}(\epsilon^n)$, one needs to consider
vertices at order ${\cal O}(\epsilon^{n-2\ell})$ for
diagrams of $\ell$-loops. 

In the definition of a physical observable, there
is a certain number of infrared factors involved. 
For instance, the spin polarizability is defined
from a Compton amplitude which contains two factors
of photon fields, one factor of photon momentum 
$q$, and two powers of photon energy $\nu$. 
Therefore, when Feynman diagrams are calculated at 
${\cal O}(\epsilon^n)$, the contribution to the
spin polarizability has infrared power $\epsilon^{n-5}$. 
Since the spin-dependent polarizability starts
at ${\cal O}(\epsilon^3)$, we expect $\gamma$
to start at order $\epsilon^{-2}$. In other words,
we have the following expansion in infrared parameters,
\begin{equation}
      \gamma = c_{-2} \epsilon^{-2} + 
          c_{-1} \epsilon^{-1} + c_0 + c_1\epsilon  + ...\,\,\,\,\, .
\end{equation}
The calculation presented in Ref. \cite{bernard} corresponds to 
$c_{-2}$. In this paper, we are interested in $c_{-1}$. 
Since the delta contribution
at this order involves new unknown parameters, 
we mainly concentrate on how to get the nucleon 
contribution to $c_{-1}$ in chiral perturbation theory. 

We immediately rule out the tree nucleon contribution
to $c_{-1}$ because the tree cannot possibly contain
the infrared parameter $m_\pi$ in the denominator. 
For this reason, the tree nucleon diagram starts
to contribute to the spin polarizability only 
at ${\cal O}(p^5)$. Therefore we need to 
consider only the one-loop nucleon diagram at
${\cal O}(p^4)$. As we have discussed above, this
involves vertices at ${\cal O}(p)$ and 
${\cal O}(p^2)$. The Feynman rules in a general
electromagnetic gauge can 
be easily derived from the lagrangian, 
for instance, in Ref. \cite{meissner}.  

\begin{figure}
\label{fig1}
\epsfig{figure=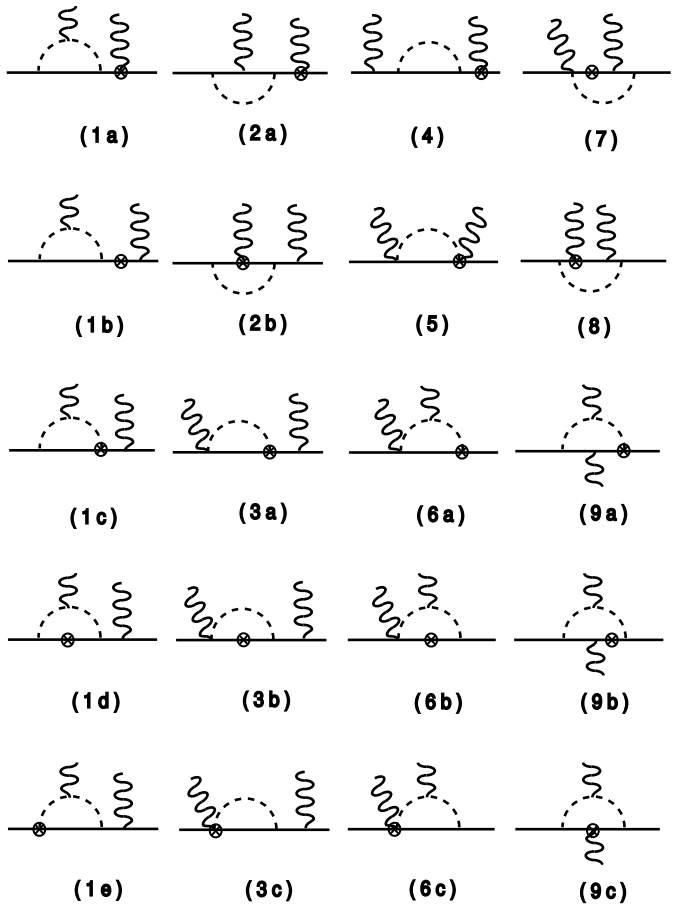,height=12cm}
\vspace{0.3in}
\caption{The diagrams that contribute to 
$S_{1,2}(\nu,Q^2)$ at NLO in heavy baryon $\chi$PT.  Obviously,
the diagrams from hermiticity and crossing must 
also be included.  The cross
indicates an insertion from ${\cal L}^{(2)}$.} 
\end{figure}   

$S_1(\nu)$ is calculated as a function of the photon
energy $\nu$. At ${\cal O}(p^4)$, there are 20 
nonzero Feynman diagrams and their close relatives. 
These diagrams are shown in Fig. 1, where the cross
in each diagram represents an insertion from ${\cal L}^{(2)}$. We find
\begin{eqnarray}
   S_1^{{\cal O}(p^4)}(\nu)& =& {g_A^2m_\pi^4\over 192\pi^2f_\pi^2
         M\nu^2\sqrt{m_\pi^2-\nu^2}}
     \left[\pi\left(1+{9\over 2}\kappa_S +{25\over 2}
    \kappa_V - (7+ {7\over 2}\kappa_S - {9\over 2}\kappa_V)
    \tau^3\right)
    \right. \nonumber \\  && 
    \times \sqrt{1-\left({\nu\over m_\pi}\right)^2} 
   - 8\left(-2+2\kappa_V-2(2+\kappa_S)\tau^3
    + \left({\nu\over m_\pi}\right)^2(1-4\kappa_V+(2+\kappa_S)\tau^3)
    \right. \nonumber \\ &&
    \left. + \left({\nu\over m_\pi}\right)^4((4+2\kappa_V)
      +(2+\kappa_S)\tau^3)\right) 
      \arccos\left(-{\nu\over m_\pi}\right)
    \nonumber \\   && \left.
    +~ 12\tau^3\kappa_S{\nu\over m_\pi}\sqrt{1-\left(
       {\nu\over m_\pi}\right)^2}
     \arccos^{2}\left(-{\nu\over m_\pi}\right)\right] 
     + (\nu\rightarrow -\nu) \ , 
\label{result}
\end{eqnarray}
where $\kappa_S=-0.120 $ and $\kappa_V = 3.706$
are the experimental values of the isoscalar and isovector anomalous magnetic
momentum of the nucleon, respectively.
$S_1(\nu)$ has a pole at $\nu=0$ 
which comes from the nucleon 
elastic contribution. According to the discussion
in Ref. \cite{jiosborne}, this pole can be 
subtracted away, with the remainder representing
the inelastic contribution (denoted with an overline).
A plot of ${\overline S_1}^{{\cal O}(p^4)}(\nu)$ is shown in 
Fig. 2. 

\begin{figure}
\label{fig2}
\epsfig{figure=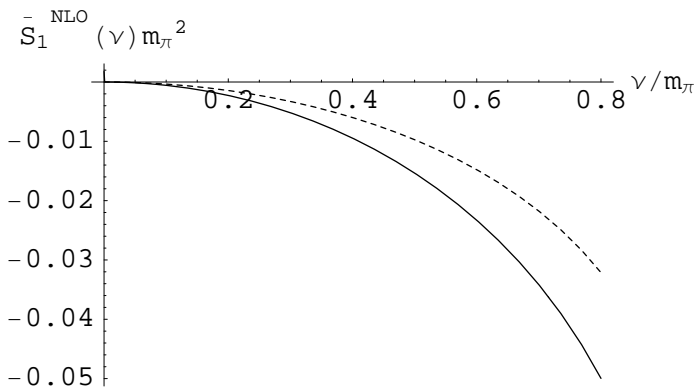,height=6cm}
\vspace{0.3in}
\caption{${\overline S_1}^{{\cal O}(p^4)}(\nu)$ as a 
function of $\nu$ below the pion threshold.  This contribution 
dominates the leading order for both proton (solid) and 
neutron (dashed) targets. We see the beginning of the 
characteristic cusp at threshold.}
\end{figure}   
  
As a check of our result, we set
$\nu=0$ : 
\begin{equation}
  {\overline S_1}^{{\cal O}(p^4)}(0) = 
   {g_A^2 m_\pi \over 8\pi f_\pi^2 M}
      \left(\kappa_S \tau^3 + \kappa_V\right) \ . 
\label{low}
\end{equation}
This is exactly the result required by the low-energy
theorem, which states that to all orders in perturbation theory
\begin{equation}
   {\overline S_1}(0) = - {\kappa^2\over M^2} \ . 
\end{equation}
Using the known result that $\kappa_V = \kappa_V^0 - 
g_A^2 m_\pi M /(4\pi f_\pi^2) + {\cal O}(p^2)$, where $\kappa_V^0$
is the isovector anomalous magnetic moment in
the chiral limit, we see that ${\overline S_1}$ 
must be Eq. (\ref{low}) at order $O(p^4)$.

Expanding Eq. (\ref{result}) as a series
in $\nu^2$, we get the contribution to $\gamma$ 
\begin{equation}
  \gamma^{{\cal O}(p^4)} 
= - {\alpha_{\rm em}g_A^2[(15+3\kappa_V) + (6+\kappa_S)\tau^3]
     \over 192\pi f_\pi^2 m_\pi M} \ . 
\end{equation}
Numerical results can be obtained by substituting
in the physical values of the parameters : 
\begin{eqnarray}
  \gamma_p^{{\cal O}(p^4)} &=& -8.2 
    \cdot 10^{-4}~ {\rm fm}^4  \nonumber \\
  \gamma_n^{{\cal O}(p^4)}  &=& -5.2 \cdot 10^{-4} ~
{\rm fm}^4\,\,\, .
\end{eqnarray}
One notices immediately that these numbers are 
somewhat larger than
those from order ${\cal O}(p^3)$.  
This certainly calls the convergence of 
the chiral expansion into question. On the
other hand, one may argue that the chiral terms 
at even and odd orders are fundamentally 
different as exemplified by the difference
in the overall $\pi$ factors; it is possible that the next
terms are much smaller than the terms considered here.

Finally, we come to the $\Delta$ contribution at
${\cal O}(\epsilon^4)$. 
In contrast to the nucleon intermediate states, 
there is a contribution at tree order because we
treat the mass difference $\Delta$ as small. 
Indeed, the mass difference $\Delta$ 
appears in the denominator of the delta 
propagator. The tree contributions come with 
one insertion of the vertices from
${\cal L}^{(2)}$ and ${\cal L}^{(3)}$. 
Since ${\cal L}^{(3)}$
contains a number of unknown parameters, we 
will not consider its contribution here.

One can also consider the one-loop 
contribution with intermediate delta states. 
In this case, only one insertion from 
${\cal L}^{(2)}$ is needed.
However, considering the fact that the 
delta-loop contribution is small at 
${\cal O}(p^3)$, we suspect that the contribution at
this level is also small. Therefore, it 
has been neglected. 

To summarize, we have calculated the spin 
polarizability of the nucleon at ${\cal O}(p^4)$. 
The result indicates a large correction to 
the ${\cal O}(p^3)$ value. 

\acknowledgements
This work is supported in part by funds provided by the
U.S.  Department of Energy (D.O.E.) under cooperative agreement
DOE-FG02-93ER-40762.


\begin{references}
\frenchspacing

\bibitem{review}
A. M. Bernstein and B. R. Holstein, Proceedings
on Chiral Dynamics: Theory and Experiment, Cambridge, MA, 
USA, 1994. 

\bibitem{f2}
S. D. Drell and A. C. Hearn, Phys. Rev. Lett. {\bf 16}, 
908 (1966). 

\bibitem{bernard}
V. Bernard, N. Kaiser, J. Kambor, and Ulf-G. Meissner, {\bf B388}, 315 
(1992). 
 
\bibitem{heavybaryon}
E. Jenkins and A. V. Manohar, Phys. Lett. {\bf B255} (1991) 558. 

\bibitem{hemmert}
T. R. Hemmert, B. R. Holstein, J. Kambor, and 
G. Kn\"ochlein, Phys. Rev. D {\bf 57}, 5746 (1998). 
  
\bibitem{meissner}
V. Bernard, N. Kaiser, and Ulf-G. Meissner, 
Int. J. Mod. Phys. E4, 193 (1995). 

\bibitem{jiosborne}
X. Ji and J. Osborne, hep-ph/9905410.

\nonfrenchspacing
\end{references}
\end{document}